%% file: polexprx_arxiv.tex
\documentclass[conference,final]{IEEEtran}
\usepackage{amsmath,amssymb}
\usepackage{bbm}
\usepackage{graphicx}
\include{macros}

\begin{document}
\title{Asymptotic Moments for Interference Mitigation in Correlated Fading Channels}
\author{
\IEEEauthorblockN{
Jakob Hoydis\IEEEauthorrefmark{1}\IEEEauthorrefmark{2}, M\'{e}rouane Debbah\IEEEauthorrefmark{2}, and Mari Kobayashi\IEEEauthorrefmark{1}} 
\IEEEauthorblockA{\IEEEauthorrefmark{1}Department of Telecommunications, Sup\'{e}lec, 91192 Gif-sur-Yvette, France}
\IEEEauthorblockA{\IEEEauthorrefmark{2}Alcatel-Lucent Chair on Flexible Radio, Sup\'{e}lec, 91192 Gif-sur-Yvette, France\\
\{jakob.hoydis, merouane.debbah, mari.kobayashi\}@supelec.fr}
}
\maketitle

\begin{abstract}
We consider a certain class of large random matrices, composed of independent column vectors with zero mean and different covariance matrices, and derive asymptotically tight deterministic approximations of their moments. This random matrix model arises in several wireless communication systems of recent interest, such as distributed antenna systems or large antenna arrays. Computing the linear minimum mean square error (LMMSE) detector in such systems requires the inversion of a large covariance matrix which becomes prohibitively complex as the number of antennas and users grows. We apply the derived moment results to the design of a low-complexity polynomial expansion detector which approximates the matrix inverse by a matrix polynomial and study its asymptotic performance. Simulation results corroborate the analysis and evaluate the performance for finite system dimensions.
\end{abstract}

\section{Introduction}
Distributed antenna systems and large antenna arrays have recently attained significant research interest \cite{gesbert2010,Marzetta2009}. Both are considered as promising solutions to counter intercell interference and to increase the spectral efficiency of current cellular networks. Since these techniques rely in essence on a significant increase of the number of \emph{coordinated antennas}, the computational complexity of the joint precoding/detection of the transmitted/received signals grows. This calls for low-complexity solutions. In this paper, we address this need by assessing the performance of a polynomial expansion detector \cite{Moshavi1996} adapted to the following general channel model.

Consider a discrete-time $N\times K$ multiple-input multiple-output (MIMO) channel with output vector $\yv\in\CC^N$:
\begin{align}\label{eq:chn}
 \yv  = \Hm\xv + \nv
\end{align}
where $\xv=[x_1,\dots,x_K]\tp$ is the complex channel input vector satisfying $\expect{\xv\xv\htp}=\Id_K$, $\Hm=[\hv_1\cdots\hv_K]\in\CC^{N\times K}$ is the random channel matrix and $\nv\sim\Cc\Nc(\zerov,\sigma^2\Id_N)$ is a vector of additive noise. The $j$th column $\hv_j\in\CC^N$ of $\Hm$ is modeled as 
\begin{align}\label{eq:columns}
 \hv_j=\frac{1}{\sqrt{K}}\Rm_j\wv_j,\quad j=1,\dots,K
\end{align}
where $\Rm_j\in\CC^{N\times N}$ is a deterministic matrix and the elements of $\wv_j\in\CC^N$ are independent and identically distributed (i.i.d.) random variables with zero mean, unit variance and finite eighth moment. This channel model captures different types of wireless communication systems and generalizes several well-known channel models as discussed below:

\textit{Distributed Antenna Systems:} 
Let $\Rm_j=\diag\LB r_{1j},\dots,r_{Nj}\RB$ with elements $r_{ij}=\sqrt{p_j}/d_{ij}^{\beta/2}$, where $d_{ij}$ is the (normalized) distance between transmitter $j$ and receive antenna $i$, $\beta$ is the path loss exponent and $p_j$ is the transmit power of transmitter $j$. This model is suitable for distributed antenna systems \cite{gesbert2010} where each transmitter sees a different path loss to each of the receive antennas since $d_{1j},\dots,d_{Nj}$ are different. \vspace{5pt}

\textit{Large-scale MIMO:} Assume a receiver equipped with a very large antenna array ($N\gg 1$) as in \cite{Marzetta2009}. Unless the antenna spacing is sufficiently large, it is likely that the received signals at different receive antennas are correlated. Our model allows to assign a different correlation matrix $\Rm_j$ to each transmitter.\vspace{5pt}

\textit{MIMO Multiple Access Channel (MAC):}
Consider a MIMO MAC from $M$ transmitters equipped with $K_m$, $m=1,\dots,M$, antennas to a receiver with $N$ antennas. Each point-to-point link has a different transmit and receive correlation matrix \cite{couillet10}: 
$$\yv=\sum_{m=1}^M\Phim_{\text{R},m}^{\frac12}\Wm_m\Phim_{\text{T},m}^{\frac12}\xv_m+\nv$$
where $\Phim_{\text{R},1},\dots,\Phim_{\text{R},M}\in\CC^{N\times N}$ are deterministic correlation matrices, $\Phim_{\text{T},1}\in\CC^{K_1\times K_1},\dots,\Phim_{\text{T},M}\in\CC^{K_M\times K_M}$ are nonnegative diagonal matrices, $\Wm_1\in\CC^{N\times K_1},\dots,\Wm_M\in\CC^{N\times K_M}$ are random channel matrices with i.i.d.\ entries with zero mean and variance $1/K$, and $\xv_1\in\CC^{K_1},\dots,\xv_M\in\CC^{K_M}$ are the transmit vectors. Let $\sum_{m=1}^M K_m=K$. Setting $\Rm_{j}=\Phim_{\text{R},m}^{1/2}[\Phim_{\text{T},m}^{1/2}]_{ii}$ for $j\in\{1+\sum_{l=1}^{m-1}K_l,\dots,\sum_{l=1}^{m}K_l\}$ and $i=j-\sum_{l=1}^{m-1}K_l$, we fall back to the model in \eqref{eq:columns}. \vspace{10pt}

In the sequel, we will study the asymptotic behavior of the moments $\mu_n$ of the matrix $\Bm\defines\Hm\Hm\htp$, defined as 
\begin{align}
 \mu_n \defines  \frac1N\trace\Bm^n,\quad n=0,1,2,\dots
\end{align}
under the assumption that $N$ and $K$ grow infinitely large at the same speed. In particular, we will derive deterministic approximations $\overline{\mu}_n$ of $\mu_n$ , such that $\mu_n -\overline{\mu}_n\to 0 $ almost surely, for $N,K\to\infty$. This result can be used, for example, to compute low-complexity approximations of the matrix inverse $(\Bm+\sigma^2\Id_N)^{-1}$. The computation of this matrix arises in many practical applications, such as for linear multiuser detectors and beamforming strategies. We will focus exemplary on the linear minimum mean square error (LMMSE) detector.

The LMMSE estimate $\hat{\xv}$ of $\xv$, assuming perfect knowledge of $\Hm$  at the receiver, is given as \cite{verdu_book}
\begin{align}\label{eqn:MMSE}
 \hat{\xv} &= \Hm\htp(\Bm+\sigma^2\Id_N)^{-1}\yv .
\end{align}
The computational complexity of this estimate is of order $\Oc(r^2)$ \cite{cottatellucci05}, where $r=\min(N,K)$. A reduced complexity estimate can be obtained by approximating the matrix inverse in \eqref{eqn:MMSE} by the following matrix polynomial \cite{Moshavi1996}
\begin{align}\label{eq:pol_exp}
(\Bm+\sigma^2\Id_N)^{-1} \approx \sum_{l=0}^{L-1}w_l \Bm^{l}
\end{align}
for some coefficients $w_l$, where the filter rank $L\le r$ is chosen according to the allowable complexity. For a given transmitter $k$, the above \emph{polynomial expansion detector} can be seen as a projection of $\yv$ on the $L$th Krylov subspace associated to the pair $(\Bm,\hv_k)$, i.e., the subspace of $\CC^N$ spanned by the vectors $\{\hv_k,\Bm\hv_k,\dots,\Bm^{L-1}\hv_k\}$, and a weighting of the joint projections by the coefficients $w_l$. 
Depending on $L$, the polynomial expansion detector achieves a performance between the matched filter ($L=1$) and the LMMSE detector ($L=r$) \cite{Moshavi1996} and allows, consequently, to trade-off performance for complexity. Moreover, \eqref{eq:pol_exp} allows for an efficient multistage implementation \cite{Moshavi1996,mueller01,cottatellucci05}, where each stage $l$ consists of a matched filter $\Hm\htp$ and subsequent ``re-spreading'' by the matrix $\Hm$.  In \cite{loubaton03}, it was shown that the signal-to-interference-plus-noise ratio (SINR) at the filter output converges in certain cases exponentially in the filter rank $L$ to the SINR output of the LMMSE detector. Thus, $L$ does not need to scale with the system size to achieve close to optimal performance \cite{honig01}.

The optimal weight vector $\wv=\LSB w_0\cdots w_{L-1}\RSB\tp$ can be chosen to minimize the mean square error of the estimated vector $\hat{\xv}$, i.e.,
\begin{align}\label{eqn:opt_w}
 \wv = \arg\min_{\uv=[u_0,\dots,u_{L-1}]\tp} \expect{\left\lVert\xv - \Hm\htp\sum_{l=0}^{L-1}u_l \Bm^l\yv\right\rVert_2^2}\ .
\end{align}
The solution to this optimization problem is given as \cite{Moshavi1996}
\begin{align}\label{eq:opt_weights}
 \wv\ =\ \Phim^{-1}\varphiv
\end{align}
where $\Phim\in\RR_+^{L\times L}$ and $\varphiv\in\RR_+^L$ are defined as
\begin{align}\label{eq:coeff1}
 \LSB\Phim\RSB_{ij} &\ =\ \mu_{i+j} + \sigma^2\mu_{i+j-1}\\\nonumber
\LSB\varphiv\RSB_i &\ =\ \mu_{i}.
\end{align}

The computation of the weight vector $\wv$ requires the calculation of the moments $\mu_1,\dots,\mu_{2L}$ which is still computational expensive for large $L$. 
However, under the assumption that the dimensions of $\Hm$ grow infinitely large, it was shown for several random matrix models (e.g.\ \cite{mueller01,honig01,li04}) that the moments $\mu_n$ can be closely approximated by their asymptotic counterparts $\bar{\mu}_n$. These are independent of a particular realization of $\Hm$  and can be calculated based on the statistical properties of the channel matrix. If these properties change on a much slower timescale than the fast-fading channel fluctuations, the weight vector $\wv$ can be precomputed using $\bar{\mu}_n$ instead of $\mu_n$. Thus, the detector complexity depends only on the complexity of the projection on the Krylov subspace which is of order $\Oc(r)$ \cite{cottatellucci05}. 
\newpage

Multistage or reduced-rank multiuser detectors were mainly considered in the context of code-division multiple-access (CDMA) systems as low-complexity solutions to the joint detection of a large number of user terminals with long spreading sequences \cite{Moshavi1996}. The asymptotic (universal) weight design was first studied in 
\cite{mueller01} for the equal transmit power case and then extended to more involved models, such as different transmit powers \cite{honig01,li01}, multi-path fading \cite{li04} and random unitary spreading sequences \cite{hachem04}. These results were then put on a common ground in \cite{cottatellucci05} which compares different types of linear multistage detectors in terms of their complexity and asymptotic performance. Recently, also multistage detectors for asynchronous CDMA systems were considered in \cite{cottatellucci10}.

The asymptotic results in the above works are based on the almost sure (a.s.) convergence of the empirical spectral distribution (e.s.d.) of the matrix $\Bm$ to a compactly supported limit distribution. This limit distribution is in general given implicitly by its Stieltjes transform which can be computed based on the statistical properties of the underlying random matrix model. The asymptotic moments are then obtained by writing the Stieltjes transform as a moment generating function \cite[Theorem 2.3]{couilletRMT} and relying on combinatorial arguments \cite{li04} or free probability theory \cite{hachem04}.

The technique used in this work is different in two aspects. First, we do not require the existence of a limiting eigenvalue distribution of the matrix $\Bm$. Instead, we provide for each pair $(N,K)$ a deterministic approximation $\overline{\mu}_n$ of the moments $\mu_n$ which becomes arbitrarily tight as $N,K\to\infty$. Second, the moments are derived through iterated differentiation of the Stieltjes transform and can be computed by simple recursive equations. This is in contrast to \cite{li04} which requires an exhaustive search over complicated sets of indices. Hence, our results are more practical from an implementation perspective. Moreover, the asymptotic moments of the random matrix model \eqref{eq:columns} have not been considered in the literature before.\vspace{3pt}

The paper is structured as follows: Section~\ref{sec:def} contains definitions and related results. The asymptotic moments of $\Bm$ are derived in Section~\ref{sec:mom} and the performance of the polynomial expansion receiver is studied in Section~\ref{sec:perf}. Numerical results are provided in Section~\ref{sec:num}. Section~\ref{sec:con} concludes the paper.

\section{Related results}\label{sec:def}
We need the following definitions and related results. Denote by ``$\Rightarrow$'' and ``$\xrightarrow[]{\text{a.s.}}$'' weak and almost sure convergence.\vspace{3pt}

\begin{definition}[Empirical spectral distribution]\label{def:esd} Let $\Am\in\CC^{N\times N}$ be a Hermitian matrix with eigenvalues $\lambda_1,\dots,\lambda_N$. Denote $F^{\Am}$ the e.s.d. of $\Am$, defined as
$$ F^{\Am}(x)= \frac1N\sum_{i=1}^N \mathbbm{1}(\lambda_i\le x) .$$ 
\end{definition}\vspace{3pt}

\begin{definition}[Stieltjes transform]\label{def:ST}
Let $F$ be a real measurable function over $\RR$ with support $\supp(F)$. For $z\in\CC\setminus\supp(F)$, the Stieltjes transform $m_{F}(z)$ of $F$ is defined as
$$ m_{F}(z) =  \int_{-\infty}^\infty\frac1{\lambda-z}dF(\lambda) .$$
Denote by $\Sc$ the class of functions $f$ analytic over $\CC\setminus\RR_+$, such that, for $z\in\CC_+$, $f\in\CC_+$, $zf\in\CC_+$ and $\lim_{y\to\infty}-{\bf i}yf({\bf i}y)<\infty$. Such functions are known to be Stieltjes transforms of finite measures supported by $\RR_+$ \cite[Theorem 2.2]{couilletRMT}.
\end{definition}\vspace{5pt}

\begin{theorem}[{\cite[Theorem 1]{wagner2011}}]\label{th:detequ}
 Let $\Dm\in\CC^{N\times N}$ be a Hermitian non-negative definite matrix and assume that $\Dm$ and the matrices $\Rm_j$, $j=1,\dots,K$, have uniformly bounded spectral norms (with respect to $N$). Let $N,K\to\infty$, such that $0<\lim\inf\frac KN \le \lim\sup \frac KN < \infty$. Then, for any $z\in\CC\setminus\RR_+$,
$$
 \frac1N\trace\Dm\LB\Bm-z\Id_N\RB^{-1} - \frac1N\trace\Dm\Tm(z) \xrightarrow[]{\text{a.s.}} 0
$$
where $\Tm(z)\in\CC^{N\times N}$ is defined as
\begin{align}\label{eq:T}
 \Tm(z) \defines \LB\frac1K\sum_{j=1}^K\frac{\Rm_j\Rm_j\htp}{1+\delta_j(z)}  -z\Id_N\RB^{-1}
\end{align}
and the following set of $K$ implicit equations
$$
 \delta_j(z)  =  \frac1K \trace\Rm_j\Rm_j\htp\Tm(z),\quad j=1,\dots,K
$$
admits a unique solution $(\delta_1(z),\dots,\delta_K(z))\in\Sc^K$. Moreover, denote by $F$ the distribution function whose Stieltjes transform is given by $m(z)=\frac1N\trace\Tm(z)$. Then, almost surely,
\begin{align*}
F^{\Bm} - F \Rightarrow 0 .
\end{align*}
\end{theorem}

\section{Asymptotic Moments}\label{sec:mom}
In this section, we state our main results. The proofs of Theorems~\ref{th:mom} and \ref{th:convergence} are provided in the appendix.

\vspace{5pt}
\begin{theorem}\label{th:mom}
 Let $F$ be the distribution function as defined in Theorem~\ref{th:detequ} and denote by $\overline{\mu}_0,\overline{\mu}_1,\dots$ the successive moments of $F$, defined as $\overline{\mu}_n\defines\int_{0}^\infty\lambda^ndF(\lambda)$. These moments can be calculated as
$$
 \overline{\mu}_{n}  =   \frac{(-1)^{n}}{n!}\frac1N\trace\Tm_n
$$
where $\Tm_n$ is defined recursively by the following set of equations for $n\ge0$:
\begin{align*}
\Tm_{n+1} =& \sum_{i=0}^n\sum_{j=0}^i\binom{n}{i}\binom{i}{j}\Tm_{n-i}\Qm_{i-j+1}\Tm_j \\
\Qm_{n+1} =& \frac{n+1}K\sum_{k=1}^Kf_{k,n}\Rm_k\Rm_k\htp\\
f_{k,n+1} =& \sum_{i=0}^n\sum_{j=0}^{i}\binom{n}{i}\binom{i}{j}(n-i+1)f_{k,j}f_{k,i-j}\delta_{k,n-i}\\
\delta_{k,n+1} = & \frac1K\trace\Rm_k\Rm_k\htp\Tm_{n+1}
\end{align*}
where $\Tm_0=\Id_N$, $f_{k,0} = -1$ and $\delta_{k,0} = \frac1K\trace\Rm_k\Rm_k\htp\ \forall k$. 
\end{theorem}\vspace{5pt}

\begin{remark}
While Theorem~\ref{th:mom} allows to compute the moments $\overline{\mu}_n$ of $F$, it does not imply the a.s.\ convergence of $\mu_n$ and $\overline{\mu}_n$ in general. Theorem~\ref{th:convergence} provides some sufficient conditions for which this convergence holds.
\end{remark}\vspace{5pt}

\begin{remark}
 Although difficult to show analytically, one can verify numerically that Theorem~\ref{th:mom} coincides with \cite[Theorem 1]{li04} for $\Rm_j=\diag(r_{1j},\dots,r_{Nj})$, $j=1,\dots,K$.
\end{remark}\vspace{5pt}

If the matrices $\Rm_j$ are drawn from a finite set of matrices, we get the following stronger result:

\vspace{5pt}
\begin{theorem}\label{th:convergence}
For fixed $M>0$, let $\Rc=\{\tilde{\Rm}_1,\dots,\tilde{\Rm}_M\}$ be a set of complex $N\times N$ matrices and let $\Dm\in\CC^{N\times N}$ be a non-negative definite Hermitian matrix. Assume that $\Dm$ and $\tilde{\Rm}_m,\ m=1,\dots,M$, have uniformly bounded spectral norms (with respect to $N$). Let $\Rm_j\in\Rc$ for $j=1,\dots,K$. Assume $N,K\to\infty$, such that $0<\lim\inf\frac KN \le \lim\sup \frac KN < \infty$. Then, for $n=0,1,2,\dots$,
\begin{align*}
 \frac1N\trace\Dm\Bm^n - \frac{(-1)^n}{n!}\frac1N\trace\Dm\Tm_n \xrightarrow[]{\text{a.s.}} 0
\end{align*}
where $\Tm_n$ is given by Theorem~\ref{th:mom}. This implies in particular,
\begin{align*}
 \mu_n - \overline{\mu}_n \xrightarrow[]{\text{a.s.}} 0.
\end{align*}
\end{theorem}\vspace{5pt}

Loosely speaking, Theorem~\ref{th:detequ} states that, for large matrix dimensions, the e.s.d.\ $F^{\Bm}$ of the matrix $\Bm$ can be closely approximated  by a deterministic distribution function $F$. Thus, the optimal weighting vector $\wv$ can be approximated by replacing the moments $\mu_n$ of $F^{\Bm}$ in \eqref{eq:coeff1} by the moments $\overline{\mu}_n$ of $F$. Using the result of Theorem~\ref{th:mom},  we can compute an approximate weight vector $\overline{\wv}=[\overline{w}_0\dots\overline{w}_{L-1}]$ as
\begin{align}
 \overline{\wv} = \overline{\Phim}^{-1}\overline{\varphiv}
\end{align}
where $\overline{\Phim}\in\RR_+^{L\times L}$  and $\overline{\varphiv}\in\RR_+^L$ are defined by
\begin{align}
\LSB\overline{\Phim}\RSB_{ij}  =& \overline{\mu}_{i+j}+\sigma^2\overline{\mu}_{i+j-1}\\\nonumber
\LSB\overline{\varphiv}\RSB_i  = & \overline{\mu}_{i}.
\end{align}

\section{Asymptotic Performance Analysis}\label{sec:perf}
We consider now the asymptotic performance of the polynomial expansion receiver in terms of the received SINR $\gamma_k$ for a given transmitter $k$. With weight vector $\wv$, the $k$th element $\hat{x}_k$ of the estimated vector $\hat{\xv}$ reads
\begin{align}
 \hat{x}_k = \hv_k\htp \sum_{l=0}^{L-1}w_l\Bm^l\LB\Hm\xv+\nv\RB.
\end{align}
One can easily show that the associated SINR $\gamma_k$ can be expressed as \cite[Eq. (18)]{cottatellucci05}
\begin{align}\label{eq:sinr}
 \gamma_k = \frac{\wv\tp\varphiv_k\varphiv_k\tp\wv}{\wv\tp\LB\Phim_k-\varphiv_k\varphiv_k\tp\RB\wv}
\end{align}
where $\Phim_k\in\RR_+^{L\times L}$ and $\varphiv_k\in\RR_+^{L}$ are given as
\begin{align}\label{eq:sinrdef}
 \LSB\Phim_k\RSB_{ij}&=\LSB\Bm^{i+j}\RSB_{kk} + \sigma^2\LSB\Bm^{i+j-1}\RSB_{kk}\\\nonumber
\LSB\varphiv_k\RSB_i&=\LSB\Bm^{i}\RSB_{kk}.
\end{align}
The next theorem provides a tight deterministic approximation of the terms $\LSB\Bm^{n}\RSB_{kk}=\hv_k\htp\Bm^{n-1}\hv_k$ in the asymptotic limit. 

\vspace{5pt}
\begin{theorem}\label{th:Rkk}
 Under the assumptions of Theorem~\ref{th:convergence}, the following convergence holds:
$$\LSB\Bm^{n}\RSB_{kk} - \overline{\mu}^k_n \xrightarrow[]{\text{a.s.}} 0$$
where
$$ \overline{\mu}^k_n = \sum_{i=0}^{n-1} \overline{\mu}^k_{n-i-1} \frac{(-1)^i}{i!}\frac{1}{K}\trace\Rm_k\Rm_k\htp\Tm_i,\quad n\ge1$$
and $\Tm_n$ is given by Theorem~\ref{th:mom}. The initial values of the recursion are $\overline{\mu}^k_0=1$ and $\Tm_0=\Id_N$.
\end{theorem}\vspace{5pt}
\begin{IEEEproof}[Proof of Theorem \ref{th:Rkk}]
 The proof follows the same steps as \cite[Theorem~1]{cottatellucci05} and will not be given here.
\end{IEEEproof}\vspace{5pt}

Replacing $\LSB\Bm^{n}\RSB_{kk}$  in \eqref{eq:sinrdef} by $\overline{\mu}^k_n$ and $\wv$ in \eqref{eq:sinr} by $\overline{\wv}$, we can obtain a deterministic approximation of the SINR $\gamma_k$ at the output of the polynomial expansion receiver. 

\section{Numerical Results}\label{sec:num}
Consider a MAC from $K=40$ single-antenna transmitters to a receiver with $N=100$ antennas. We use an extended version of Jake's model \cite{couillet10} for the generation of the matrices $\Rm_j$. Let $\Rm_j=\Thetam_j^{1/2}$ and $\Thetam_j\in\CC^{N\times N}$ be defined as
$$
 \LSB\Thetam_j\RSB_{kl}=\frac{1}{\phi^j_\text{max}-\phi^j_\text{min}}\int_{\phi^j_\text{min}}^{\phi^j_\text{max}}\exp\LB \frac{2\pi \bf{i}}{\lambda}d_{kl}\cos(x)\RB dx
$$
where $d_{kl}=2\lambda(k-l)$ and $\phi^j_\text{min}$, $\phi^j_\text{max}$ are drawn independently from the intervals $[-\pi,0]$ and $[0,\pi]$, respectively.
The interval $[\phi^j_\text{min},\phi^j_\text{max}]$ can be seen as the angular spread of the signal from transmitter $j$, $\lambda$ is the wave length, and $d_{kl}$ is the spacing between the receive antennas $k$ and $l$. We assume Rayleigh fading channels, i.e., $\wv_j$ in \eqref{eq:columns} are independent standard complex Gaussian vectors. The covariance matrices $\Thetam_j$ are chosen at random at the beginning and then kept fixed while we average over many realizations of the channel matrix $\Hm$. We denote by $\text{SNR}=1/\sigma^2$ the transmit signal-to-noise ratio.

\begin{figure}
\centering
 \includegraphics[width=0.48\textwidth]{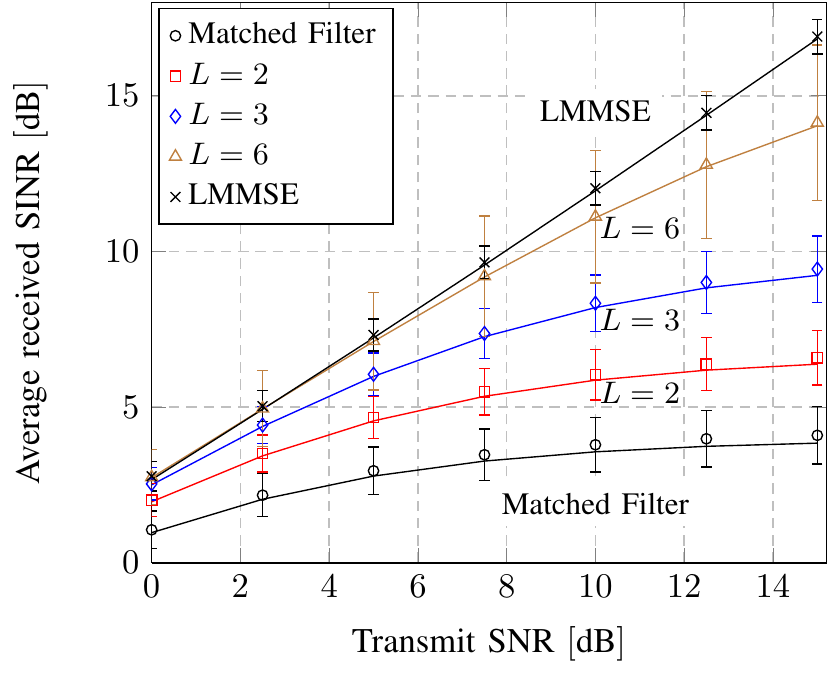}
\caption{Average received SINR versus SNR at the output of the matched filter, LMMSE detector and the polynomial expansion detector with approximate weights for different values of $L$. Markers correspond to simulation results, solid lines to the deterministic SINR approximations. Error bars indicate one standard deviation in each direction.\label{fig:sinr_vs_snr}}
\end{figure}

Fig.~\ref{fig:sinr_vs_snr} shows the average received SINR $\mathbb{E}[\gamma_k]$ of a randomly chosen transmitter as a function of the SNR for the matched filter, the LMMSE detector and the polynomial expansion detector with approximate weights for $L=\{2,3,6\}$. Markers correspond to simulation results and solid lines to the deterministic SINR approximations. The error bars indicate one standard deviation of $\gamma_k$ in each direction. Similar to \cite{hoydis2010}, the asymptotic SINR of transmitter $k$ for the LMMSE detector can be easily shown to satisfy  $$\overline{\gamma}^\text{LMMSE}_k=\frac1K\trace\Rm_k\Rm_k\htp\Tm(-1/\text{SNR})$$ where $\Tm(z)$ is given by Theorem~\ref{th:detequ}. We observe a good fit between the deterministic approximations and the simulation results for the average SINR. However, the standard deviation of the SINR increases with $L$. This is because the higher order moments converge slower to their deterministic approximations and exhibit therefore stronger fluctuations. Nevertheless, the average SINR performance of the polynomial expansion detector with $L=6$ is already close to the performance of the LMMSE detector. 

\begin{figure}
\centering
 \includegraphics[width=0.48\textwidth]{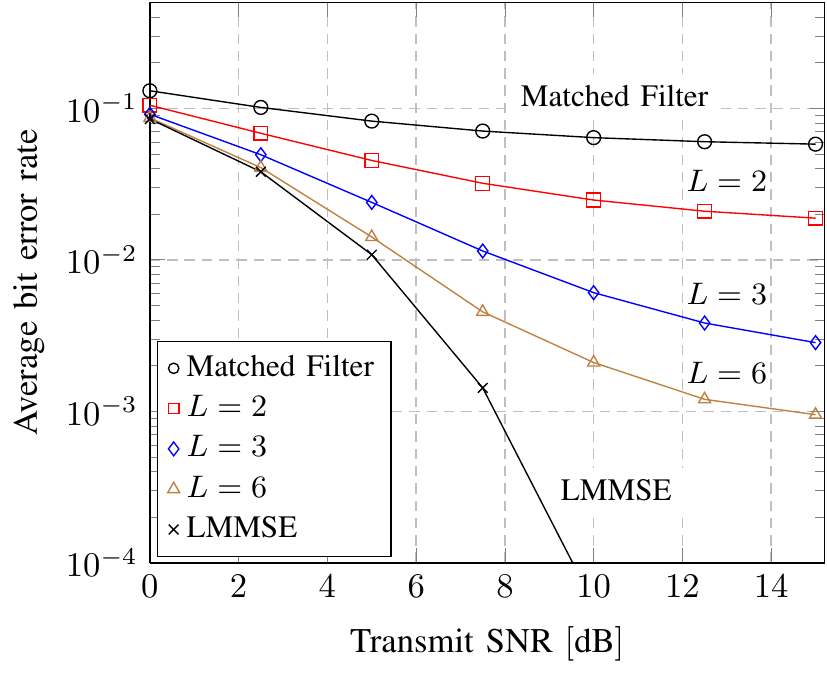}
\caption{Average theoretical bit error rate versus SNR for the matched filter, LMMSE detector and the polynomial expansion detector with approximate weights for different values of $L$. \label{fig:ber_vs_snr}}
\end{figure}

Fig.~\ref{fig:ber_vs_snr} depicts the theoretical average bit error rate (BER) over SNR for the different detectors. Assuming binary phase-shift keying (BPSK) modulation and Gaussian interference, the BER is given as $\mathbb{E}[Q(\sqrt{\gamma_k})]$ where $Q(x)$ is the Gaussian tail function. We can clearly see a performance increase of the polynomial expansion detector with $L$, although the BER saturates at high SNR. Although not explicitly shown here, one can even observe a performance decrease for large values of $L$. As mentioned before, this is due to the low accuracy of the approximate weights caused by a slow convergence of the higher-order moments to their deterministic approximations. 
  
\section{Conclusion}\label{sec:con}
We have derived asymptotically tight deterministic approximations of the moments of a certain class of large random matrices, useful for the study of distributed antenna systems and large antenna arrays. We have applied these moment results to the design of a polynomial expansion detector
which significantly reduces the computational complexity of multiuser detection compared to the LMMSE detector. Moreover, we have derived an explicit expression of the asymptotic SINR at the output of this detector and verified its accuracy and performance for finite system dimensions by simulations. 

\appendix
\begin{IEEEproof}[Proof of Theorem \ref{th:mom}]
From Definition~\ref{def:ST}, it is easy to see that the moments $\overline{\mu}_n$ of the distribution function $F$ can be obtained through successive differentiation of the function $\frac1z m(-\frac1z)$, i.e., 
\begin{align}\nonumber
 \overline{\mu}_n&=\frac{(-1)^n}{n!}\frac{d^n}{dz^n}\left.\LB\frac1z m\LB-\frac1z\RB\RB\right|_{z=0}\\\nonumber
&=\left.\frac{(-1)^n}{n!}\frac{d^n}{dz^n}\LB\int\frac{1}{z\lambda+1} dF(\lambda)\RB\right|_{z=0}\\\nonumber
&=\left.\frac{(-1)^n}{n!}\int\frac{d^n}{dz^n}\LB\frac{1}{z\lambda+1}\RB dF(\lambda)\right|_{z=0}\\\nonumber
&=\int\lambda^n dF(\lambda)
\end{align}
where we could exchange the order of differentiation and integration since $1/(z\lambda+ 1)dF(\lambda)$ is analytic for $z\ge 0$. Consider now the following function for $z\ge0$:
\begin{align*}
 \eta(z) = \frac1z m\LB-\frac1z\RB
\end{align*}
and denote $\eta_n(z)$ its $n$th derivative with respect to $z$. 
From Theorem~\ref{th:detequ}, we have
\begin{align*}
 \eta(z)&= \frac1z m\LB-\frac1z\RB\\
&= \frac1N\trace\LB z \frac1K\sum_{j=1}^K\frac{\Rm_j\Rm_j\htp}{1+\delta_j\LB-\frac{1}{z}\RB}+\Id_N\RB^{-1}\\
&= \frac1N\trace\LB z \frac1K\sum_{j=1}^K\frac{\Rm_j\Rm_j\htp}{1+z\delta_{j,0}(z)}+\Id_N\RB^{-1}\\
&= \frac1N\trace \Tm_0(z)
\end{align*}
where 
\begin{align*}
 \Tm_0(z) \defines \LB z\frac1K\sum_{j=1}^K\frac{\Rm_j\Rm_j\htp}{1+z\delta_{j,0}(z)} + \Id_N\RB^{-1}
\end{align*}
and $\LB\delta_{1,0}(z),\dots,\delta_{K,0}(z)\RB\in\RR_+^K$ is the unique solution to the $K$ implicit equations:
\begin{align*}
 \delta_{j,0}(z)= \frac1K \trace\Rm_j\Rm_j\htp\Tm_0(z),\quad j=1,\dots,K.
\end{align*}

Denoting $\Tm_n(z)=\frac{d^n \Tm_0(z)}{dz^n}$, we have
\begin{align*}
  \eta_n(z) =  \frac1N\trace\Tm_n(z).
\end{align*}
In order to find the derivatives $\Tm_n(z)$, we need the following additional definitions. For $k\in\{1,\dots,K\}$, let
\begin{align*}
g_{k,0}(z) & = z\delta_{k,0}(z)\\
f_{k,0}(z) &= -\frac{1}{1+g_{k,0}(z)}\\
t_{k,0}(z) &= zf_{k,0}(z)
\end{align*}
and denote $\delta_{k,n}(z)$, $g_{k,n}(z)$, $f_{k,n}(z)$, and $t_{k,n}(z)$ their $n$th derivatives, respectively. Furthermore, let
\begin{align*}
 \Qm_0(z) = \frac1K\sum_{k=1}^Kt_{k,0}(z)\Rm_k\Rm_k\htp 
\end{align*}
and denote $\Qm_n(z)=\frac{d^n \Qm_0(z)}{dz^n}$. 
We continue by writing
\begin{align}
 \Tm_1(z) = \Tm_0(z)\underbrace{\Qm_1(z)\Tm_0(z)}_{\defines\Gm_0(z)}.
\end{align}
From the Leibniz-rule for the $n$th derivative of the product of two functions\footnote{For two functions $u(x)$ and $v(x)$, $\frac{d^n \LB u(x)v(x)\RB}{dx^n}=\sum_{i=0}^n\binom ni \frac{d^{n-i}u(x)}{d x^{n-i}} \frac{d^iv(x)}{dx^i}$.}, we have
\begin{align*}
 \Tm_{n+1}(z) =& \sum_{i=0}^n\binom{n}{i}\Tm_{n-i}(z)\Gm_i(z),\quad n\ge 0\\
 \Gm_{n}(z) =& \sum_{i=0}^n\binom{n}{i}\Qm_{n-i+1}(z)\Tm_i(z),\quad n\ge 0
\end{align*}
where  $\Gm_n(z)=\frac{d^n \Gm_0(z)}{dz^n}$. Replacing the last equation in the second last yields
\begin{align}\label{eq:Tn}
 \Tm_{n+1}(z) = \sum_{i=0}^n\sum_{j=0}^i\binom{n}{i}\binom{i}{j}\Tm_{n-i}(z)\Qm_{i-j+1}(z)\Tm_j(z).
\end{align}
Straight-forward differentiation of $\Qm_0(z)$ leads to
\begin{align}\label{eq:Qn}
 \Qm_n(z) =  \frac1K\sum_{k=1}^Kt_{k,n}(x)\Rm_k\Rm_k\htp,\quad n\ge0.
\end{align}
The last step is to find explicit expressions of $t_{k,n}(z)$. From the Leibniz-rule, we have
\begin{align*}
 t_{k,n}(z) = nf_{k,n-1}(z)+zf_{k,n}(z)\ ,\qquad n\ge0.
\end{align*}
Consider now $f_{k,1}(z)$ the first derivative of $f_{k,0}(z)$:
\begin{align*}
 f_{k,1}(z) = \frac{g_{k,1}(z)}{\LB1+g_{k,0}\RB^2}=\underbrace{f^2_{k,0}(z)}_{\defines r_{k,0}(z)}g_{k,1}(z).
\end{align*}
The higher order derivatives are calculated as
\begin{align*}
 f_{k,n+1}(z) = \sum_{i=0}^n\binom{n}{i}r_{k,i}(z)g_{k,n-i+1}(z)
\end{align*}
where
\begin{align*}
 r_{k,n}(z) = \sum_{i=0}^{n}\binom{n}{i}f_{k,i}(zk)f_{k,n-i}(z).
\end{align*}
Combining the last two equations yields
\begin{align}\label{eq:fn}
  f_{k,n+1}(z) = \sum_{i=0}^n\sum_{j=0}^{i}\binom{n}{i}\binom{i}{j}f_{k,j}(z)f_{k,i-j}(z)g_{k,n-i+1}(z)
\end{align}
where $g_{k,n}(z)$ can be easily calculated as
\begin{align*}
 g_{k,n}(z) = n\delta_{k,n-1}(z) + z\delta_{k,n}(z)
\end{align*}
and $\delta_{k,n}(z)$ is given by
\begin{align}\label{eq:dn}
 \delta_{k,n}(z) = \frac1K\trace\Rm_k\Rm_k\htp\Tm_n(z) .
\end{align}
Since we are only interested in the case $z=0$, we will drop from now on the dependence on $z$ and write, e.g., $\Tm_n$ instead of $\Tm_n(0)$. In this case, the expressions of $g_{k,n}(z)$ and $t_{k,n}(z)$ simplify to
\begin{align*}
  g_{k,n}\ =&\ n\delta_{k,n-1}\\
  t_{k,n}\ =&\ nf_{k,n-1}.
\end{align*}
Replacing these quantities in \eqref{eq:Qn} and \eqref{eq:fn}, together with \eqref{eq:Tn} and \eqref{eq:dn} leads to the desired result. Note that $\Tm_0=\Id_N$, $f_{k,0}=-1$ and $\delta_{k,0}=\frac1K\trace{\Rm_k\Rm_k\htp}$. Moreover, $\Tm_{n+1}$ depends on $\Tm_0,\dots,\Tm_n$ and $\Qm_1,\dots,\Qm_{n+1}$. Since $\Qm_{n+1}$ depends only on $f_{k,0},\dots,f_{k,n}$ and $f_{k,n}$, $\Tm_{n+1}$ can be recursively calculated from the given initial values.
\end{IEEEproof}\vspace{5pt}

\begin{IEEEproof}[Proof of Theorem \ref{th:convergence}]
Both $\frac1N\trace\Dm\LB\Bm-z\Id_N\RB^{-1}$ and $\frac1N\trace\Dm\Tm(z)$ as defined in Theorem~\ref{th:detequ} are Stieltjes transforms of finite measures which we denote by $\pi$ and $\overline{\pi}$, respectively. Thus, Theorem~\ref{th:detequ} also implies that, almost surely,
$$\pi - \overline{\pi} \Rightarrow 0.$$
Similar to the proof of Theorem~\ref{th:mom} we can express the moments of $\pi$ and $\overline{\pi}$ as
\begin{align*}
 \int\lambda^n\pi(d\lambda)&=\frac{(-1)^n}{n!}\frac{d^n}{dz^n}\left.\LB\frac{1}{z}\frac1N\trace\Dm\LB\Bm+\frac1z\Id_N\RB^{-1}\RB\right|_{z=0}\\
&= \frac1N\trace\Dm\Bm^n
\end{align*}
and
\begin{align*}
 \int\lambda^n\overline{\pi}(d\lambda)&=\frac{(-1)^n}{n!}\frac{d^n}{dz^n}\left.\LB\frac{1}{z}\frac1N\trace\Dm\Tm(-1/z)\RB\right|_{z=0}\\
&= \frac{(-1)^n}{n!} \frac1N\trace\Dm\Tm_n.
\end{align*}
The support of $\pi$ is almost surely compact as $\Dm$ has bounded spectral norm and the spectral norm of $\Bm$ is almost surely bounded due to the following inequalities:
\begin{align*}
 \left\lVert\Bm \right\rVert &\leq \sum_{m=1}^M \left\lVert\tilde{\Rm}_m\tilde{\Rm}_m\htp \right\rVert\left\lVert\frac1K\Wm_m\Wm_m\htp \right\rVert\\
&\leq M R \sup_m\left\lVert\frac1K\Wm_m\Wm_m\htp \right\rVert\\
&\xrightarrow[]{\text{a.s.}}  MR\sup_m\frac{K_m}{K}\LB1+\sqrt{\frac{N}{K_m}}\RB^2<\infty
\end{align*}
for some $R\ge\sup_m\lVert\tilde{\Rm}_m\tilde{\Rm}_m\htp\rVert$,  $K_m\defines\sum_{j=1}^K\mathbbm{1}(\Rm_j=\tilde{\Rm}_m)$, and $\Wm_m\in\CC^{N\times K_m}$ being random matrices with i.i.d.\ elements with zero mean, unit variance and finite eighth moment. The almost sure convergence of the spectral norm in the last step follows from \cite{yin88}. 
The almost sure weak convergence of $\pi$ and $\overline{\pi}$ implies by \cite[Theorem 25.8 (ii)]{billingsley}, that
\begin{align}\label{eq:conv}
  \int f(\lambda)\pi(d\lambda) - \int f(\lambda)\overline{\pi}(d\lambda) \xrightarrow[]{\text{a.s.}}0
\end{align}
for any bounded, continuous function.
 Since the support of $\pi$ is almost surely bounded and the support of $\overline{\pi}$ can be shown to be bounded following similar steps as in 
\cite[Proof of Theorem 2, Part B]{couillet10}, the convergence in \eqref{eq:conv} also holds for any continuous function. Choosing $f(\lambda)=\lambda^n$ concludes the proof.
\end{IEEEproof}

\bibliographystyle{IEEEtran}
\bibliography{IEEEabrv,bibliography}

\end{document}

%% file: macros.tex
\newcommand{\expect}[1]{{\mathop{{\mathbb E}}}\left[#1\right]}

\newcommand{\LB}{\left(}
\newcommand{\RB}{\right)}
\newcommand{\LSB}{\left[}
\newcommand{\RSB}{\right]}
\newcommand{\htp}{^{\sf H}}
\newcommand{\tp}{^{\sf T}}

\newfont{\bbb}{msbm10 scaled 500}

\newfont{\bb}{msbm10 scaled 1100}
\newcommand{\CC}{\mbox{\bb C}}
\newcommand{\RR}{\mbox{\bb R}}

\newcommand{\hv}{{\bf h}}

\newcommand{\nv}{{\bf n}}

\newcommand{\uv}{{\bf u}}
\newcommand{\wv}{{\bf w}}

\newcommand{\xv}{{\bf x}}
\newcommand{\yv}{{\bf y}}

\newcommand{\zerov}{{\bf 0}}

\newcommand{\Am}{{\bf A}}
\newcommand{\Bm}{{\bf B}}

\newcommand{\Dm}{{\bf D}}

\newcommand{\Gm}{{\bf G}}
\newcommand{\Hm}{{\bf H}}
\newcommand{\Id}{{\bf I}}

\newcommand{\Qm}{{\bf Q}}
\newcommand{\Rm}{{\bf R}}

\newcommand{\Tm}{{\bf T}}

\newcommand{\Wm}{{\bf W}}

\newcommand{\Cc}{{\cal C}}

\newcommand{\Nc}{{\cal N}}
\newcommand{\Oc}{{\cal O}}

\newcommand{\Rc}{{\cal R}}
\newcommand{\Sc}{{\cal S}}

\newcommand{\Phim}{\hbox{\boldmath$\Phi$}}

\newcommand{\Thetam}{\hbox{\boldmath$\Theta$}}

\newcommand{\varphiv}{\hbox{\boldmath$\varphi$}}

\newcommand{\supp}{{\hbox{Supp}\,}}

\newcommand{\diag}{{\hbox{diag}}}

\newcommand{\trace}{{\hbox{tr}\,}}

\renewcommand{\arg}{{\hbox{arg}}}

\newcommand{\defines}{{\,\,\stackrel{\scriptscriptstyle \bigtriangleup}{=}\,\,}}

\newtheorem{theorem}{Theorem}
\newtheorem{definition}{Definition}

\newtheorem{remark}{Remark}[section]